\documentclass{aa}
\usepackage{graphics,astron,textcomp}
\title{ISO observations of a sample of 60\,$\mu$m peaker galaxies}

\author{R.~J.~Laureijs\inst{1}
	\and D.~Watson\inst{2}
	\and L.~Metcalfe\inst{1}
	\and B.~McBreen\inst{2}
	\and B.~O'Halloran\inst{2}
	\and J.~Clavel\inst{1}
	\and K.~Leech\inst{1}
	\and P.~Gallais\inst{1}
	\and P.~Barr\inst{1}
	\and M.~Delaney\inst{3}
	\and L.~Hanlon\inst{2}
	\and F.~Quilligan\inst{2}
}

\thesaurus{3(
		09.04.1; 
		09.13.2; 
		11.09.4; 
		11.19.1; 
		11.19.3; 
		13.09.1)
}

\institute{  ISO Data Centre, Astrophysics Division, Space Science Department of
	   ESA, Villafranca del Castillo, P.O. Box 50727, E-28080 Madrid, Spain.
	\and Department of Physics, University College, Dublin 4, Ireland
	\and Stockholm Observatory, SE-133 36 Saltsj\"{o}baden, Sweden
}

\offprints{dwatson@bermuda.ucd.ie}

\date{Received / Accepted}

\begin{document}
\maketitle


\begin{abstract}
The sample of IRAS galaxies with spectral energy distributions that peak
near 60\,$\mu$m are called Sixty Micron Peakers (SMPs or 60PKs). Their
generally peculiar and amorphous morphologies, hot dust and lack of a cirrus
component have been interpreted as being indicative of a recent
interaction/merger event.
Mid-infrared spectra of eight SMPs, obtained with ISOPHOT-S in the
$\sim$2--11\,$\mu$m band  are presented.  Four of the observed sources
are H\,II region-like (H2) galaxies, three are Seyfert 2 and one is
unclassified.
Emission attributed to Polycyclic Aromatic Hydrocarbons (PAHs) at
6.2\,$\mu$m, 7.7\,$\mu$m and 8.6\,$\mu$m is ubiquitous in the
spectra.
The PHOT-S spectrum of the H2 galaxy \object{IRAS\,23446+1519} exhibits a
bright 11.04\,$\mu$m line and an 8.6\,$\mu$m feature of comparable size to
its 7.7\,$\mu$m feature.
[S~IV] emission at 10.5\,$\mu$m was detected in three of four H2 galaxies and in
one Seyfert 2 galaxy.
The ratio of the 7.7\,$\mu$m PAH feature to the continuum at 7.7\,$\mu$m
(PAH L/C) divides the eight SMPs at a ratio greater than 0.8 for H2 and less
than 0.8 for Seyfert galaxies.
An anti-correlation between PAH L/C and the ratio of the continuum flux
at 5.9\,$\mu$m to the flux at 60\,$\mu$m is found, similar to that found in
ultraluminous infrared galaxies.
Silicate absorption at approximately 9.7\,$\mu$m was
observed in the Seyfert 2 galaxy, \object{IRAS\,04385-0828} and in
\object{IRAS\,03344-2103}.
The previously unclassified SMP galaxy
\object{IRAS\,03344-2103} is probably a Seyfert~2.
\end{abstract}

\section{Introduction}

IRAS galaxies with spectral energy distributions peaking near 60\,$\mu$m
\cite{1993AJ....106.1743V} are known as Sixty Micron Peakers (SMPs or
60PKs).  Vader et al. \cite*{1993AJ....106.1743V} constructed the sample
using the IRAS flux ratios $f_{60}/f_{100}>1$ and $1<f_{60}/f_{25}<4$ to
select extragalactic sources with galactic latitude $|b|>10\degr$ from the
second version of the IRAS Point Source Catalogue.
There are only 51
galaxies in the sample, constituting about 2\% of the space density of
60\,$\mu$m-selected galaxies in the range L$_{60\,\rm\mu m}$ =
10$^{9}$--10$^{12}$L$_{\sun}$ that have been identified out to a redshift of
0.2.

All SMPs are strong emission-line galaxies
\cite{1993AJ....106.1743V}. The optical line ratio criteria of Veilleux \&
Osterbrock \cite*{1987ApJS...63..295V} have been used to classify SMPs as
H\,II region-like (H2) or Seyfert (Sy) galaxies.  Of the 41 SMPs that have
been classified, 23 are Sy2, 5 are Sy1 and 13 are H2.  Approximately 60\% of
SMPs are therefore Seyfert galaxies.  This fraction is far higher than the
percentage of Seyferts found in samples of ultraluminous IRAS galaxies
\cite{1994AJ....107...35H}.  There are very few Sy1 galaxies in the full
sample and none of the galaxies presented here is a Sy1.  Only 4 of a sample
of 45 SMPs have spiral optical morphologies, the majority being amorphous and
peculiar \cite{1994AJ....107...35H}. The paucity of spiral galaxies is
another notable feature of SMPs, since IRAS galaxies are typically
spirals.
The luminosity range of SMPs of about three orders of magnitude,
spans the classes from dwarf galaxies to giant ellipticals.


Near-infrared (NIR) brightness profiles of H2
galaxies are well modelled by a single r$^{1/4}$ fit but an additional nuclear
point source component is required to fit the majority of Seyfert galaxy
profiles \cite{1995AJ....110...87H}.  The radio continuum and H$\alpha$ emission is
compact, indicating that the FIR radiation is also emitted from a small
volume \cite{1998MNRAS.300.1111H}.  The compact region has an extent of a
few kpc which is comparable in size to typical narrow emission-line regions
in AGN.  The central region of SMPs therefore appears to be dominated by
strong non-thermal or starburst emission.  This emission is heavily
obscured at optical wavelengths. 
A short-lived phase of central activity, caused by a
recent interaction/merger, accounts for the morphology, the brightness near
60\,$\mu$m and the strong non-thermal or starburst source
\cite{1995AJ....110...87H}.

The $f_{60}/f_{100}>1$ criterion preferentially selects galaxies with warmer
or more centrally located dust.
The `cirrus' component which is the dominant
contributor to the 100\,$\mu$m flux in spiral galaxies must therefore be
weak or absent in SMPs \cite{1993AJ....106.1743V}.  Very small grains (VSGs)
of carbonaceous material contribute strongly to the flux between 25 and 60\,$\mu$m
\cite{1990A&A...237..215D,Laureijs:1998} and SMPs may be dominated
by emission from VSGs.
There is still some debate as to the actual carriers of the CH and CC aromatic
bonds that produce emission features that dominate in the 3--13\,$\mu$m spectral
region of spiral galaxies \cite{Lu:1999} and Seyferts
\cite{Clavel:1999}.  It is commonly assumed that the emission
is produced by polycyclic aromatic hydrocarbons (PAHs) but other models
exist such as the `Coal Model' \cite{1989A&A...217..204P,Guillois:1998}.  Laboratory and
theoretical results based on PAH mixtures fit the observed spectra consistently
\cite{Salama:1998,1998A&A...339..194B,Moutou:1998,1996JPCA..100.2819L}.  The
features at 3.3, 6.2, 7.7, 8.6 and 11.3\,$\mu$m are therefore referred to as
`PAH emission' hereafter.  PAH emission is complex
\cite{Peeters:1999,1998A&A...339..194B,1996JPCA..100.2819L} and
varies spatially within galaxies
\cite{Tielens:1999,Moorwood:1999}.  The 2--11\,$\mu$m spectra of SMPs are
expected to be largely dominated by the ubiquitous PAH emission. 
Their spectra should however be modified by direct emission from non-thermal
or starburst components and from hot dust (VSGs) \cite{1990A&A...237..215D}. 
The presence of high energy photons will also modify the spectrum indirectly
by exciting and destroying different PAHs preferentially depending on their
composition \cite{1998ApJ...493L.109U,1996A&A...315L.289R}.

Sect.~2 details observations of eight SMPs with the ISOPHOT
photopolarimeter \cite{1996A&A...315L..64L} on board the Infrared Space
Observatory (ISO) \cite{1996A&A...315L..27K}.  Results of the observations
and a discussion are presented in Sect.~3.  The separation of H2 and Sy
galaxies is discussed in Sect.~4.  Conclusions are in Sect.~5.  In this
paper, $H_{0}$ = 75\,km\,s$^{-1}$Mpc$^{-1}$ and $q_{0}$ = 0.5 are adopted.


\section{Observations and Data Reduction}

Eight of the brightest SMPs visible to ISO were selected for observation
with ISOPHOT-S being roughly representative of the full SMP sample.  Four
of the galaxies are H2, three are Sy2, and the classification of one is uncertain.
Observations were carried out between December 1997 and
February 1998. Table~\protect\ref{obs} lists the R.A., Dec. and the
redshift of the galaxies as well as the ISO data archive number and the
date of the observation.

\begin{table*}
\begin{center}
\caption{Observations of SMPs}
\label{obs}
\begin{tabular}{@{}l c c c c c@{}}
\hline
\hline
IRAS		& R.A.		& Dec.		& z		& Archive	& Date\\ 
Galaxy		& J2000		& J2000		& 		& TDT No.		&\\
\hline
& h\, m \,\,\,\,\,\,\,\,\, s& $\degr$ \,\, \arcmin \,\,\,\,\,\,\, \arcsec&&	&	    \\
00160-0719	& 00 18 35.90	& -07 02 57.5	& 0.0180	& 74900248	& 03/12/1997\\
01475-0740	& 01 50 02.50 	& -07 25 49.3	& 0.0177	& 76300953	& 17/12/1997\\
\\
02530+0211	& 02 55 34.99 	& +02 23 12.9	& 0.0276	& 82101050	& 13/02/1998\\
03344-2103	& 03 36 39.00 	& -20 54 06.4	& 0.0058	& 81201754	& 05/02/1998\\
04385-0828	& 04 40 55.48 	& -08 23 06.6	& 0.0151	& 83301373	& 25/02/1998\\
\\
05189-2524	& 05 21 01.40 	& -25 21 45.9	& 0.0419	& 70102085	& 17/10/1997\\
08007-6600	& 08 01 09.39 	& -66 08 33.1	& 0.0412	& 72401151	& 09/11/1997\\
23446+1519	& 23 47 08.64 	& +15 36 16.1	& 0.0258	& 56500952	& 03/06/1997\\
\hline
\end{tabular}
\end{center}
\end{table*}

PHT-S consists of a dual grating spectrometer with a resolving power of
$\lambda/\Delta\lambda \sim 90$
in two wavelength bands.  Band SS covers the wavelength range
2.5--4.8\,$\mu$m, while band SL covers the range 5.8--11.8\,$\mu$m
\cite{iso-laureijs-1998esa}.  PHT-S measurements were made in chopped mode,
using either rectangular or triangular chopping with the ISOPHOT focal-plane
chopper; the mode depending on the difficulties of avoiding any nearby
sources. During the chop cycle, 512 seconds were spent on-target and 512
seconds off-target -- either in a single off-target pointing, or two 256
second off-target pointings. Chopper throw ranged from about 1--3\arcmin~
depending on the extent of the target source and the density of surrounding
field sources.
%
     The calibration of the spectrum was performed using a spectral response
     function derived from several calibration stars of different
     brightnesses observed in a mode similar to that of the
     observation of the target \cite{Acosta:1999}.
The relative photometric uncertainty of the PHT-S spectrum is better than
20\% when comparing different parts of the spectrum that are more than a few
microns apart.  The absolute photometric uncertainty is better than 30\% for
bright calibration sources \cite{Schulz:1999}.
All data processing was performed
using the ISOPHOT Interactive Analysis (PIA V7.22) system
\cite{iso-gabriel-1998esa}.  Data reduction consisted primarily of the
removal of instrumental effects such as cosmic ray glitches.  After
background subtraction was performed, flux densities for the sources were
determined.
In order to increase the signal-to-noise ratio per channel the
spectra were smoothed using:
\begin{equation}
      as_{n} =  (0.25\times a_{n-1}) + (0.5\times a_{n}) + (0.25\times a_{n+1})
\end{equation}
where $a_{n}$ is the flux in channel n and $as_{n}$ is the smoothed flux in channel n.
     Using this method, the flux is effectively spread over 2 channels in
     a way that does not change the position of the spectral peaks and
     which conserves the flux.
These fluxes were then corrected for redshift to
obtain rest-frame spectra.  The PHT-S band
fluxes were derived from these spectra.  An estimate of the continuum
was made following the methof of Lutz~et~al. \cite*{1998ApJ...505L.103L}.
A linear interpolation between the fluxes at 5.9 and 10.9\,$\mu$m was used
except for the sources \object{IRAS\,01475-0740},
\object{IRAS\,08007-6600} and \object{IRAS\,23446+1519}.  For these source, a linear
interpolation between the minimum flux values near 6\,$\mu$m and 10.5\,$\mu$m
was used instead to estimate the continuum.  Flux errors were determined by
adding in quadrature the 1\,$\sigma$ errors of all the bins in the feature.
Upper limits were derived at the 3\,$\sigma$ level of significance.


\section{Results and Discussion}

\begin{figure*}
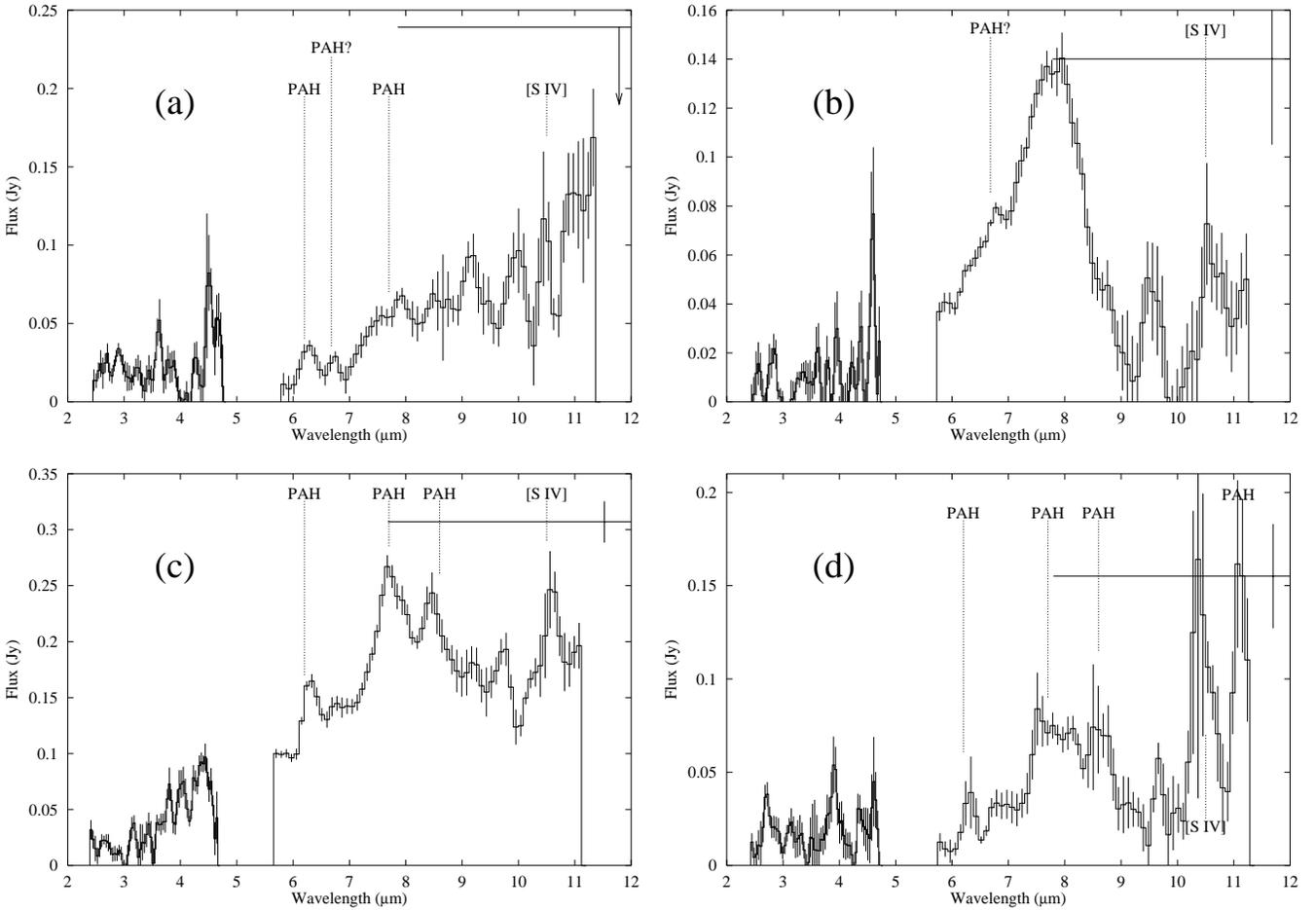

	\resizebox{\textwidth}{!}{\includegraphics*{H1884.f1a}\includegraphics*{H1884.f1b}}
	\resizebox{\textwidth}{!}{\includegraphics*{H1884.f1c}\includegraphics*{H1884.f1d}}
	\caption{PHT-S spectra of the H2 galaxies:
		 {\bf a} \object{IRAS\,00160-0719}, {\bf b} \object{IRAS\,02530+0211},
		 {\bf c} \object{IRAS\,08007-6600}, {\bf d} \object{IRAS\,23446+1519}.
		 The IRAS 12\,$\mu$m flux or upper limit is plotted for comparison.}
	\label{H2s}
\end{figure*}

\begin{figure*}
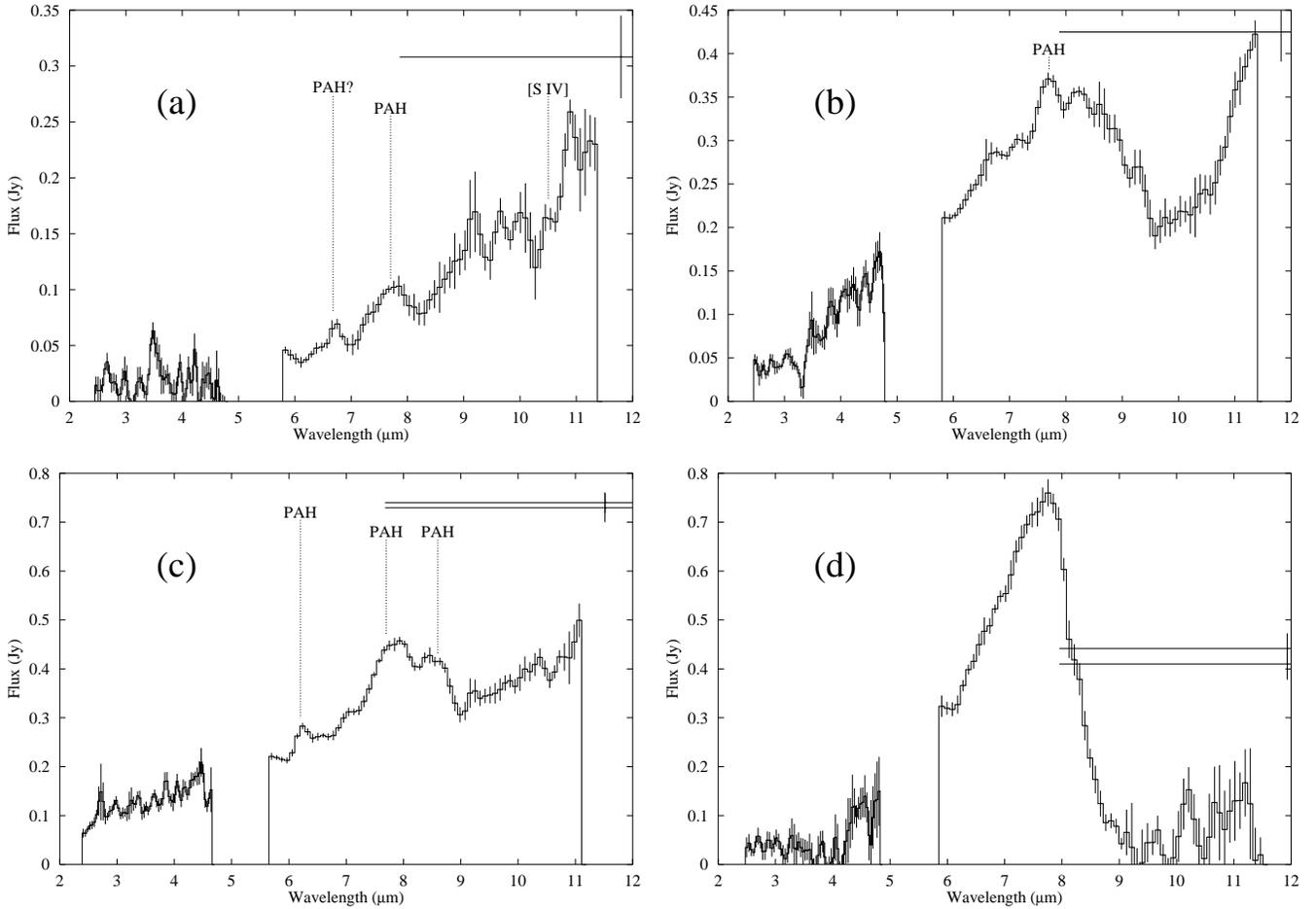

	\resizebox{\textwidth}{!}{\includegraphics*{H1884.f2a}\includegraphics*{H1884.f2b}}
	\resizebox{\textwidth}{!}{\includegraphics*{H1884.f2c}\includegraphics*{H1884.f2d}}
	\caption{PHT-S spectra of the three Sy2 galaxies: {\bf a}
		 \object{IRAS\,01475-0740}, {\bf b} \object{IRAS\,04385-0828} and
		 {\bf c} \object{IRAS\,05189-2524}, and the one unclassified
		 galaxy: {\bf d} \object{IRAS\,03344-2103}.
		 IRAS 12\,$\mu$m fluxes are plotted for comparison.}
	\label{Sy2s}
\end{figure*}

Spectra of eight SMP galaxies are presented in Figs.~\ref{H2s} and
\ref{Sy2s}. The results of these observations are summarised in
Tables~\protect\ref{fluxes} and \protect\ref{lines}.
Table~\protect\ref{fluxes} lists ISOPHOT and IRAS fluxes, ratios of the IRAS
60\,$\mu$m to the PHT-SL flux, the equivalent width (EW) of the 7.7\,$\mu$m
PAH feature and the PAH to continuum ratios at 7.7\,$\mu$m (PAH L/C), as
well as the optical spectral classification of the galaxies.  It is clear
from Table~\ref{fluxes} that the ratio of IRAS 60\,$\mu$m flux to the PHT-SL
flux is generally higher in the H2 galaxies than it is in the Seyferts. The
EW is more uncertain in galaxies with broad PAH features where the
spectrum is heavily absorbed by silicate (e.g. Fig.~\ref{Sy2s}d). The
7.7\,$\mu$m PAH EWs are generally larger in the H2 galaxies.  The ratio of
the height of the 7.7\,$\mu$m feature above the continuum to the continuum
level at 7.7\,$\mu$m is referred to as PAH L/C (Table~\ref{fluxes}). The
continuum was obtained from a linear interpolation between points near
5.9\,$\mu$m and 10.9\,$\mu$m as described in the previous section.
This ratio is a measure of the relative importance of PAH in the total emission
from the galaxy, and has been shown to be a good discriminator between starburst
and AGN in a sample of Ultraluminous Infrared Galaxies (ULIRGs)
\cite{1998ApJ...505L.103L,1998ApJ...498..579G}.  Applying this ratio
to SMPs shows that it can discriminate between H2 and Seyfert
galaxies (Table~\ref{fluxes}).

Table~\ref{lines} lists the line fluxes, and an
estimate of the continuum at the centre of the lines.  The luminosities of
the 7.7\,$\mu$m PAH feature are also listed in Table~\ref{lines}.  The
7.7\,$\mu$m luminosity gives a better estimate of the amount of PAH in the
galaxy than the other PAH features between 5 and 11\,$\mu$m, which are more
prone to dust-extinction \cite{Rigopoulou:1999}. The luminosity of this feature
appears to be independent of spectral type (Table~\ref{lines} and Clavel et
al. 1999).

\begin{table*}
\begin{center}
\caption{ISOPHOT Fluxes with IRAS fluxes listed for comparison. 
	 f$_{60}$/f$_{\mathrm{PHT}}$ is the IRAS 60\,$\mu$m flux divided by
	 the PHT-SL flux.  The spectral classification of the galaxies is
	 adopted from Vader et al. \protect\cite*{1993AJ....106.1743V}}
\label{fluxes}
\begin{tabular}{@{}l c c c c c c c c c c@{}}
\hline
\hline
IRAS 		& \multicolumn{2}{c}{ISO Fluxes (mJy)}& \multicolumn{4}{c}{IRAS\,Fluxes (Jy)} & FIR/MIR & EW	($\mu$m)	&PAH L/C & Spec.\\
GALAXY		& PHT-SS	& PHT-SL	& 12\,$\mu$m	& 25\,$\mu$m	& 60\,$\mu$m	& 100\,$\mu$m	& f$_{60}$/f$_{\mathrm{PHT}}$& (7.7\,$\mu$m) & (7.7\,$\mu$m) & Type\\
\hline
00160-0719	& $21\pm2$	& $65\pm2$	& $<$0.26	& 0.63		& 1.79		& 1.54		& 39	& 4.7	& 1.9	& H2\\
01475-0740	& $16\pm2$	& $119\pm2$	& 0.28		& 0.83		& 1.1		& $<$1.05	& 12	& 0.6	& 0.2	& Sy2\\
\\
02530+0211	& $11\pm2$	& $58\pm2$	& $<$0.25	& 0.81		& 2.77		& 1.79		& 45	& 2.6	& 1.6	& H2\\
03344-2103	& $50\pm6$	& $276\pm6$	& 0.41		& 1.91		& 7.23		& 5.96		& 22	& (0.8)$\dagger$	& 0.8	& ?\\
04385-0828	& $84\pm3$	& $289\pm2$	& 0.45		& 1.67		& 2.95		& 2.14		& 11	& 0.1	& 0.3	& Sy2\\
\\
05189-2524	& $131\pm4$	& $359\pm2$	& 0.76		& 3.52		& 13.94		& 11.68		& 45	& 0.3	& 0.4	& Sy2\\
08007-6600	& $40\pm1$	& $177\pm2$	& 0.27		& 1.11		& 3.7		& 3.19		& 24	& 0.5	& 0.9	& H2\\
23446+1519	& $17\pm2$	& $56\pm3$	& $<$0.25	& 1.22		& 4.2		& 3.54		& 100	& 8.0	& 4.4	& H2\\
\hline
$\dagger$ Tentative
\end{tabular}
\end{center}
\end{table*}

\begin{table*}
\begin{center}
\caption{Emission line fluxes or upper limits ($10^{-15}$ W\,m$^{-2}$) and
	 estimates of the continuum(C) at the centre of the line ($10^{-15}$
	 W\,m$^{-2}$\,$\mu$m$^{-1}$) for the eight sources observed with ISOPHOT.}
\label{lines}
\begin{tabular}{@{}l |c c|c c|c c c|c c|c c@{}}
\hline
\hline
Lines		& PAH		& C	&PAH		& C	& PAH	& Luminosity	& C	& PAH		& C	& [S~IV]	& C\\
$\lambda$ ($\mu$m)& 6.2	&	& 6.67		&	& 7.7	& $\times 10^{34}$\,W&	& 8.6		&	& 10.5		&\\
\hline
00160-0719	& 2.2$\pm$0.4	& 1.2	& $<$0.9	& 1.3	& 18$\pm$2	& 11	& 1.4	& 4.6$\pm$1.2	& 1.5	& 1.0$\pm$0.4	& 1.5\\
01475-0740	& $<$0.9	& 3.0	& 0.9$\pm$0.2	& 3.2	& 15$\pm$2	&  9	& 3.4	& $<$6.5	& 3.5	& 0.7$\pm$0.2	& 3.4\\
&&&&&&&&&&\\
02530+0211	& $<$1.9	& 3.0	& 0.9$\pm$0.03	& 2.7	& 59$\pm$2	& 86	& 2.3	& $<$3.5	& 2.0	& $<$0.6	& 1.5\\
03344-2103	& $<$10.8	& 22.5	& $<$13.7	& 18.8	& 299$\pm$5	& 19	& 12.1	& $<$6.3	& 7.7	& $<$1.5	& 2.5\\
04385-0828	& $<$3.4	& 16.6	& 1.5$\pm$0.2	& 14.8	& 79$\pm$2	& 35	& 12.4	& 15$\pm$1	& 10.2	& $<$0.2	& 7.5\\
&&&&&&&&&&\\
05189-2524	& 2.8$\pm$0.4	& 18.6	& $<$0.4	& 17.3	& 91$\pm$2	& 309	& 15.2	& 13.9$\pm$0.8	& 13.6	& $<$0.6	& 11.1\\
08007-6600	& 5.8$\pm$0.3	& 7.5	& $<$1.3	& 6.7	& 89$\pm$2	& 296	& 5.6	& 18$\pm$1	& 4.7	& 1.8$\pm$0.4	& 3.5\\
23446+1519	& 2.4$\pm$0.7	& 0.6	& 0.6$\pm$0.1	& 0.5	& 41$\pm$2	& 55	& 0.5	& 9.6$\pm$1.3	& 0.4	& 2.7$\pm$1.0	& 0.3\\
\hline
\end{tabular}
\end{center}
\end{table*}

The emission feature at $\sim$6.7\,$\mu$m (Table~\ref{lines}) exists in the
spectra of two H2 and one Sy2 galaxy.  It is labelled `PAH?' in
Figs.~\ref{H2s} and \ref{Sy2s} and is identified with the weak PAH feature at
6.66\,$\mu$m \cite{Peeters:1999}.

Emission features  between 9\,$\mu$m and 11\,$\mu$m exist in the spectra of 3 H2
galaxies;  \object{IRAS\,00160-0719}, \object{IRAS\,02530+0211} and 
\object{IRAS\,23446+1519} (Figs.~\ref{H2s}b,c and d) and in one Sy2
\object{IRAS\,01475-0740} (Fig. 2a) and may be attributable to
PAH.  Another possiblility however, is that these narrow peaks may be
emission features from crystalline silicates (Watson~et~al. in preparation).
Crystalline silicates have already been detected in solar system comets
\cite{1994ApJ...425..274H,1997Sci...275.1904C},
interplanetary dust particles \cite{Bradley:1998}, the disks surrounding
young stars \cite{1998A&A...332L..25M,1998Ap&SS.255...25W} and in the
outflows of evolved stars \cite{Waters:1998}.  Crystalline silicate can be
produced in a condensation sequence such as that around oxygen-rich AGB
stars or by annealing amorphous silicate at temperatures near 1000\,K
\cite{1999Nat...401..563M} or perhaps by a low-temperature annealing process
\cite{1999Nat...401..563M}.
For spectra with peaks near 9.1\,$\mu$m, assuming they are produced by
hypersthene at 300\,K, the mass of crystalline silicate required to produce
this emission is $\sim$1\,M$_{\sun}$ in \object{IRAS\,00160-0719} and
$\sim$1.6\,M$_{\sun}$ in \object{IRAS\,01475-0740}.
Spectral observations at longer wavelengths should also reveal features due
to crystalline silicate emission, further assisting identification of these
minerals.  Observations of these longer wavelength features are important in
studies of stellar sources \cite{Waters:1998}.

Though the PHT-SL band extends to 11.8\,$\mu$m, the PAH emission at
$\sim$11.3\,$\mu$m is difficult to detect in these spectra due to the
redshift of the galaxies. It can be difficult to distinguish absorption at
$\sim$9.7\,$\mu$m due to silicate, from an artefact caused by broad PAH
emission at $\sim$8\,$\mu$m and 11.3\,$\mu$m
\cite{Roche:1989}.
Silicate absorption is probably present in the spectra of
\object{IRAS\,04385-0828} and \object{IRAS\,03344-2103}
(Figs.~\ref{Sy2s}b and \ref{Sy2s}d respectively).  The shape of the trough
at 9.7\,$\mu$m, the clearly rising continuum in the PHT-SS band, and the
IRAS 12\,$\mu$m fluxes all provide strong evidence that the absorption
features in these two spectra are not artefacts. Noise dominates PHT-SS
spectra with fluxes less than $\sim0.05$\,Jy. It is clear however that
there is a strong signal in the long wavelength end of the PHT-SS spectra
of \object{IRAS\,04385-0828} and \object{IRAS\,03344-2103}.



\subsection{H2 Galaxies}

The spectra of the four H2 galaxies (Fig.~\ref{H2s}) are all dominated by
emission features associated with PAHs between 6 and 9\,$\mu$m, though
\object{IRAS\,08007-6600} (Fig.~\ref{H2s}c) clearly has a strong MIR
continuum.  The PAH emission in \object{IRAS\,02530+0211} appears to be a blend
of the standard 7.7 and 8.6\,$\mu$m features in one broad peak, but
absorption by silicate cannot be ruled out in this galaxy.
\object{IRAS\,23446+1519} has an unusual spectrum with strong line emission
near 11.0\,$\mu$m (Fig.~\ref{H2s}d).  This line has been detected in H\,II
regions (at 11.04\,$\mu$m) and attributed to PAHs.  Roelfsema et al.
\cite*{1996A&A...315L.289R} have interpreted the existence of this feature
in conjunction with an unusually strong 8.6\,$\mu$m band (comparable to the
7.7\,$\mu$m feature) and a 7.8\,$\mu$m shoulder on the 7.7\,$\mu$m feature
as being indicative of the presence of non-compact PAHs.  All these
characteristics are present in the spectrum of \object{IRAS\,23446+1519}. 
In older systems with PAH emission, the less stable non-compact PAHs will
generally have been destroyed, but in a non-equilibrium situation, the
non-compact PAHs can exist and modify the observed spectrum.  Non-compact
PAHs are probably present in \object{IRAS\,23446+1519}.  It is interesting
to note that a very similar spectrum has been observed from the Wolf-Rayet
(WR) galaxy \object{NGC~1741} (McBreen~et~al. in preparation).

The broad 9.7\,$\mu$m silicate absorption feature is more common in AGN than
in H2 type galaxies \cite{Roche:1989}.  Evidence of this feature is
present in the spectra of two galaxies, neither of which is H2 type.  [S~IV]
emission at 10.5\,$\mu$m was detected in three of the four H2 galaxies, but in only
one Sy2 (see Table~\ref{lines}).  Photons of a few eV are sufficient to
generate standard PAH emission spectra at wavelengths shorter than 9\,$\mu$m
\cite{1998ApJ...493L.109U}, but this is not the case with other lines
emitted in the MIR.  The presence of [S~IV] emission implies a flux of hard
photons in the region in which it is produced, assuming photo-ionisation. 
The more luminous MIR continuum in Seyferts would tend to diminish the
signal-to-noise ratio of the [S~IV] line and may explain the greater prevalence of
this line in the H2 galaxies.  WR stars are energetic enough to
produce [S~IV] emission and it should be noted that this line has already
been observed in the WR galaxies \object{Haro~3}
\cite{1996A&A...315L.105M}, \object{NGC~7714} \cite{Halloran:1999} and
\object{NGC~5253} \cite{1999MNRAS.304..654C}.  Some SMPs are also
classified as WR galaxies.  They are \object{NGC~5253},
\object{II\,Zw\,40}, \object{Mrk\,1210}, \object{Tol\,1924-416} and possibly
\object{Tol\,1345-419} \cite{1999A&AS..136...35S}.  A search for the optical
signature of WRs in other SMPs could reveal new WR galaxies.

\subsection{Sy2s}

In general the three Sy2 galaxies (Figs.~\ref{Sy2s}a, b and c) have similar
7.7\,$\mu$m PAH luminosities (Table~\ref{lines}), but smaller
7.7\,$\mu$m PAH EW and L/C values than the four H2 galaxies
(Table~\ref{fluxes}). It is clear from Fig.~\ref{Sy2s} that the Sy2s have
significant continuum emission thus explaining their low PAH EW and L/C
values.  Continuum emission from a central source above that detected in H2
galaxies is necessary to explain the NIR brightness profiles of Sy2 SMPs
\cite{1995AJ....110...87H}. This implies that some of the brighter continuum
observed in these Sy2 spectra could be, directly or indirectly due to the
Seyfert nucleus.  The spectrum of \object{IRAS\,03344-2103}
(Fig.~\ref{Sy2s}d) has a very large silicate absorption trough near
9.7\,$\mu$m making it difficult to identify lines beyond $\sim$8\,\textmu
m and making the EW of 7.7\,$\mu$m feature difficult to determine.

There are clear discriminators between H2 and Sy2 galaxies in the
mid-infrared.  They are PAH L/C, EW and the IRAS 60\,$\mu$m to the PHT-SL
flux ratio (see Table~\ref{fluxes} and Fig.~\ref{H2_S2}) and are discussed
in the next section.


\section{Separation of Starbursts and AGN}

In the standard model, Sy1s and Sy2s differ in their orientation to the line
of sight \cite{1995PASP..107..803U}.  The Seyfert nucleus is surrounded by a
dusty torus that obscures the broad emission-line region (BLR) in Sy2s (which
are viewed edge-on).


PAH emission is produced outside the torus and is independent of the nucleus
\cite{Clavel:1999}.  In the well-studied Sy2 \object{Circinus galaxy},
Moorwood et al. \cite*{Moorwood:1999} manage explicitly to exclude the
nucleus as the source of excitation for the PAH emission using ISOCAM-CVF.
Clavel et al. \cite*{Clavel:1999} use the EW of the 7.7\,$\mu$m feature as a
discriminator between Sy1 and Sy2.  Laurent et al. \cite*{Laurent:1999} have
proposed a similar diagnostic, using the ratio of the PAH 6.2\,$\mu$m band
to the 5.1 to 6.7\,$\mu$m continuum to distinguish AGN from starburst
galaxies.  These are in essence a similar measure to the PAH L/C at
7.7\,$\mu$m since all depend on the facts that (a) PAH luminosities are
independent of the active nucleus \cite{Clavel:1999} and (b) the MIR
continuum is stronger in Sy1 than Sy2 and brighter in AGN as a whole than in
starburst galaxies.  Laurent et al. \cite*{Laurent:1999} extend this
diagnostic tool to assess the relative contributions of components from AGN,
H\,II and photo-dissociation regions by introducing the ratio of the warm to
the hot continuum. This ratio was not obtainable in the PHT-S waveband
because it does not extend to 15\,$\mu$m.


  In Sy1s the BLR and the inner wall of the torus are directly visible, 
  making the MIR continuum brighter 
  in these galaxies. In Sy2s seen fully edge-on, our line of sight 
  to the inner wall of the torus is blocked and the MIR continuum is
  suppressed \cite{Clavel:1999}. In the
  intermediate situation of Sy2 viewed at grazing incidence,
  one has a direct view of the inner wall of the torus but a reflected
  view of the BLR.  This corresponds to Sy2s where the MIR continuum
  is strong and broad lines are observed only in polarised light
  \cite{Clavel:1999,1996MNRAS.280..579H,1997Natur.385..700H}.
  \object{IRAS\,05189-2524} falls into this category, since it
  it has polarised broad-lines \cite{1996MNRAS.281.1206Y} and a strong
  MIR continuum (Fig.~\ref{Sy2s}c).  The 7.7\,$\mu$m PAH EW of \object{IRAS\,05189-2524}
  is 0.3\,$\mu$m falling in the range observed by Clavel et al. \cite*{Clavel:1999} for
  this sub-class of Sy2. \object{IRAS\,04385-0828} has a 7.7\,$\mu$m
  EW of just 0.1\,$\mu$m (Table~\ref{fluxes}) and is therefore another
  good candidate for possessing polarised broad lines.
Unfortunately, though it was observed in polarised
light, the signal-to-noise ratio was not sufficiently good to determine the
EW of the lines in the polarised spectrum \cite{1996MNRAS.281.1206Y}.

Dudley \cite*{1999MNRAS.307..553D} finds the 1--5\,$\mu$m SED of 
\object{IRAS\,05189-2524} to be similar to that of the infrared-bright
quasar \object{IRAS\,13349+2438} \cite{1986ApJ...308L...1B}.  He therefore
proposes that hot dust heated by an AGN is responsible for the 1--5\,$\mu$m
continuum in both sources.  It is clear however that two components are
required to fit the 1--100\,$\mu$m SED in \object{IRAS\,05189-2524}.  Dudley
\cite*{1999MNRAS.307..553D} therefore further suggests that the
8--100\,$\mu$m emission is better explained by emission arising from star
formation than from an AGN. The observation of \object{IRAS\,05189-2524}
presented here shows no evidence of a spectral break between 2.5\,$\mu$m and
11\,$\mu$m.  The very large IRAS 12\,$\mu$m flux for this source indicates
that such a spectral break could lie near 12\,$\mu$m (Fig.~\ref{Sy2s}c).  It
is clear from these results that \object{IRAS\,05189-2524} is a Seyfert and
that the hot inner wall of the torus is largely responsible for the
2.5\,$\mu$m--11\,$\mu$m continuum.  Recent PHT-S observations of
\object{IRAS\,13349+2438} \cite{Watson:2000} while showing PAH emission,
also suggest that the inner wall of the torus is the source of most of the
2.5\,$\mu$m--11\,$\mu$m continuum in this quasar.  The similarity in the
spectra of \object{IRAS\,05189-2524} and \object{IRAS\,13349+2438} may be
attributed to the suggestion that they are both AGN viewed near grazing
incidence to their tori \cite{Clavel:1999,1992ApJ...400...96W}.

\begin{figure}
	\resizebox{\columnwidth}{!}{\includegraphics*{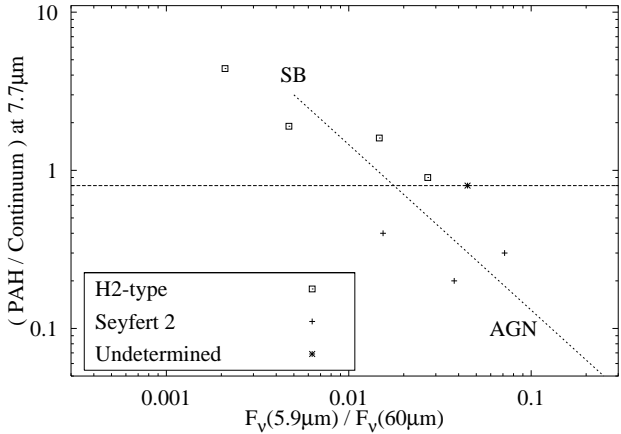}}
	\caption{Comparison of the ratio of 60\,$\mu$m flux to
		 5.9\,$\mu$m flux versus the ratio of the height
		 of the PAH feature at 7.7\,$\mu$m to the 7.7\,$\mu$m
		 continuum.  Boxes represent H2-type galaxies, plusses are
		 Seyfert 2 type galaxies, the asterisk is
		 \object{IRAS\,03344-2103}.  The dashed line at
		 PAH/Continuum = 0.8 divides H2 galaxies from Seyferts.
		 ``SB'' and ``AGN'' represent unobscured
		 starburst and AGN type emission respectively.  The dotted
		 line represents the dilution of a pure starburst spectrum
		 by a 5.9\,$\mu$m continuum as shown by Lutz et al.
		 \protect\cite*{1998ApJ...505L.103L} for a sample of
		 ULIRGs.}
        \protect\label{H2_S2}
\end{figure}

The ratio of the PHT-SL 5.9\,$\mu$m flux to the IRAS 60\,$\mu$m flux is
plotted against the PAH L/C values (Table~\ref{fluxes}) in Fig~\ref{H2_S2}.
There is an anti-correlation between the PAH L/C and the
F$_{\nu}$(5.9\,$\mu$m)/F$_{\nu}(60$\,$\mu$m) ratio indicating that warmer
SMPs are more AGN-dominated, and cooler SMPs are more starburst-like.
The
anti-correlation is significant at a probability greater than 99\% (from a 
Pearson's correlation statistic of -3.7).
The eight SMPs divide very well at PAH L/C = 0.8 (Fig.~\protect\ref{H2_S2}).
The method Lutz et al. \cite*{1998ApJ...505L.103L} applied to a sample of
ULIRGs is followed in Fig~\ref{H2_S2}.  The same anti-correlation was found for
ULIRGs as is discovered here.  Lutz et al. \cite*{1998ApJ...505L.103L} used
the criterion PAH L/C = 1 to separate AGN from starburst ULIRGs, but found this
criterion tended to classify some starbursts as AGN.

\object{IRAS\,08007-6600}, appears to be currently undergoing a merger and
possesses two distinct nuclei \cite{1994AJ....107...35H}.  It is possible
that one of the nuclei may contain a hidden non-thermal source given its
position on the Starburst-AGN plot (Fig.~\ref{H2_S2}) with PAH~L/C~=~0.9,
despite its optical classification as a H2 galaxy.  The unclassified galaxy,
\object{IRAS\,03344-2103} may also be a Seyfert since it shows evidence of
silicate absorption in its spectrum (Fig.~\ref{Sy2s}d) and has a PAH~L/C~=~0.8.

\section{Conclusions}

ISOPHOT-S spectra were obtained for a sample of eight SMPs of types H2 and Sy2.
PAH emission was detected in all the spectra.  The spectrum of
\object{IRAS\,23446+1519} shows an unusual 8.6\,$\mu$m feature with a height
comparable to the 7.7\,$\mu$m feature and exhibits a very bright
11.04\,$\mu$m PAH emission line.  The spectrum implies that non-compact PAHs are
the source of a large proportion of the PAH emission in that galaxy.  [S~IV]
emission is more prevalent in the H2 galaxies than in the
Sy2s, probably because of the brighter continuum in Sy2s. 
Silicate absorption ($\sim$9.7\,$\mu$m) was observed in
\object{IRAS\,03344-2103} and in \object{IRAS\,04385-0828}.
The results show that the H2 galaxies have PAH L/C at 7.7\,$\mu$m
$>0.8$ and the Sy2 galaxies have PAH L/C $\leq 0.8$.
The same anti-correlation exists in SMPs and ULIRGs between the PAH L/C
and the F$_{\nu}$(5.9\,$\mu$m)/F$_{\nu}(60$\,$\mu$m) ratio.
It is proposed that \object{IRAS\,08007-6600} may contain a hidden non-thermal
source and that the previously unclassified galaxy \object{IRAS\,03344-2103}
may be a Seyfert.  Observations of \object{IRAS\,04385-0828} may reveal
broad emission lines in polarised light.

While the galaxies presented here are diverse, observation of a larger
sample of SMPs in the mid-infrared could confirm the Seyfert/H2 separation
and the anti-correlation observed here.  The prevalence of [S~IV] emission
in H2 SMPs implies that a search of SMP optical spectra for the WR signature
may reveal new WR galaxies.

\bibliography{mnemonic,/users/dwatson/work/latex/refs/refs}
\end{document}